\documentclass{appolb}
\usepackage{graphicx}
\usepackage{mathtools}
\usepackage{hyperref}
\usepackage{xcolor}

\newcommand{\rmd}{{\rm d}}

\newcommand{\bx}{{\mathbf{x}}}

\newcommand{\bX}{{\mathbf{X}}}

\newcommand{\CA}{{\cal A}}

\newcommand{\CD}{{\cal D}}

\newcommand{\CI}{{\cal I}}
\newcommand{\CJ}{{\cal J}}

\newcommand{\CR}{{\cal R}}

\newcommand{\CQ}{{\cal Q}}
\newcommand{\CW}{{\cal W}}

\newcommand{\average}[1]{\left\langle #1 \right\rangle_\CD}

\newcommand{\aaverage}[1]{\left\langle #1 \right\rangle_{\CI}}

\newcommand{\inI}{{\mathrm{I}}}
\newcommand{\inII}{{\mathrm{II}}}
\newcommand{\inIII}{{\mathrm{III}}}

\begin{document}
\title{Average zero-expansion regions of the universe
\thanks{Presented at The 8th Conference of the Polish Society on Relativity }%
}
\author{Jan J. Ostrowski, Ismael Delgado Gaspar
\address{Department of Fundamental Research, National Centre for Nuclear Research, Pasteura 7, 02--093 Warsaw, Poland}
}
\maketitle
\begin{abstract}
Persistent tensions in the $\Lambda$CDM cosmological model underline the importance of tests of its basic assumptions. One such potential test arises from the fact that the surface of zero expansion around the collapsing object with spherical symmetry is strictly related to the object's mass and the value of the cosmological constant. We propose a complementary probe relating the averaged zero-expansion volume to the mass and the background cosmological Hubble parameter. Using the relativistic Zel'dovich approximation we are able to relax the spherical symmetry assumption and hence obtain a more general test of cosmological dynamics. Alternatively, our method can serve as a test of compatibility of relativistic N-body simulations and the scalar, averaged Einstein's equations with the relativistic Zel'dovich approximation serving as a closure condition.  
\end{abstract}
  
\section{Introduction}
The $\Lambda$CDM model relies on the assumption that the large-scale behaviour of the Universe is correctly described by the Friedmann-Lema\^itre-Robertson-Walker (FLRW) metric. As a consequence of applying this assumption to observations, the stress-energy tensor of the universe consists mostly of dark energy and dark matter, so far undetected directly, hypothetical sources. In the light of accumulating tensions (see e.g. \cite{tension} for a recent review) it is more important than ever to verify both the  geometry of space-time and the energy budget of the universe as described by the standard $\Lambda$CDM model. An example aiming at testing the geometrical part of this model can be found in \cite{Clarkson}. Regarding the stress-energy part, we can use the dynamics and statistics of cosmological structure formation to put some constraints on observable quantities. In particular, in \cite{Pavlidou} it was observed that each initially expanding and then collapsing structure is surrounded by a zero-expansion surface, which in the $\Lambda$CDM context has its maximum value related only to the mass and cosmological constant. This derivation was performed assuming several simplifying assumptions, e.g. the structure in question is spherical and homogeneous. Detection of the zero-expansion surface around the structure, with a radius bigger than the one dictated by the mass of the structure and the cosmological constant would therefore violate the $\Lambda$CDM model.

It is worth noting that the same result can be obtained by examining the spherically symmetric but radially inhomogeneous solution to Einstein's equations, i.e. the Lema\^itre-Tolman-Bondi (LTB) metric. The LTB metric in co-moving and synchronous coordinates reads:
\begin{equation}
\label{LTB}
\rmd s^2 = -\rmd t^2 +\frac{(R')^2}{1+2E}+R^2 \rmd \Omega^2 \;,
\end{equation}
where the comma denotes partial differentiation with respect to the radial coordinates, $E$ is a radial-dependent free function related to the spatial curvature, $R$ is the so-called areal radius and $\rmd \Omega$ is a surface element on the sphere.  
The associated Einstein's equations read:
\begin{equation}
\label{ltb-einstein}
    \dot{R}^2=2E +\frac{2GM}{R}+R^2\frac{\Lambda}{3} \;\;,\;\;4\pi G \rho  = \frac{M'}{R^2R'} \;,
\end{equation}
where $M=M(r)$ is an active gravitational mass, $\rho$ is a density and the dot stands for differentiation with respect to the proper time. Taking the time derivative of the first equation in (\ref{ltb-einstein}) and putting $\ddot{R}=0$ we find that:
\begin{equation}
\label{maxR}
    R = \left(\frac{3GM}{\Lambda}\right)^{1/3} \;,
\end{equation}
which is exactly the value obtained for the homogeneous model in \cite{Pavlidou}. Moreover, as it was shown in \cite{Bochicchio} (in a different context) it is a soft boundary for the Weierstrass function ($\CW$) given by:
\begin{equation}
    \CW = 2E+\frac{2GM}{R}+R^2 \frac{\Lambda}{3} \;,
\end{equation}
resulting from equation (\ref{ltb-einstein}) with $\dot{R}=0$ condition. 
Regardless of the method by which we obtain the value from equation (\ref{maxR}), the result is still limited to the spherical symmetry. In the consecutive section, we will present a formalism that will allow us to extend the applicability of this idea to arbitrary domains by modifying the zero-expansion condition.  
\section{Methods}
\subsection{Scalar averaging}
In order to study extended objects in the relativistic context, we will adapt the scalar averaging formalism.
Spatial averaging of the scalar parts of Einstein's equations in the synchronous comoving coordinates was developed by T. Buchert (see e.g. \cite{AveCos}).  Given an arbitrary Lagrangian domain (co-moving with the fluid) $\CD$ containing irrotational dust, the averaged Hamiltonian constraint reads: 
\begin{equation}
\label{Hamiltonian}
\frac{1}{3}\average{\theta^2} = 8\pi G \average{\rho} + \average{\sigma^2} -\frac{1}{2}\average{\mathcal{R}}+\Lambda  \;,
\end{equation}
where $\langle \rangle_{\CD}$ denotes a spatial averaging operator normalized by the domain's volume. The components of the extrinsic curvature tensor, the expansion and shear scalars, are denoted by $\theta$ and $\sigma^2$, respectively; $\rho$ is the density and $\CR$ is the spatial Ricci scalar. Combined with the non-commutation rule (for any scalar field $\CA$):
\begin{equation} 
\partial_t\langle \mathcal{A}\rangle_{\mathcal D}-\langle \partial_t \mathcal{A}\rangle_{\mathcal D}=\langle \mathcal{A}\theta\rangle_{\mathcal D}-\langle \mathcal{A}\rangle_{\mathcal D}\langle \theta\rangle_{\mathcal D} \; ,
\end{equation}
equation (\ref{Hamiltonian}) reads:
\begin{equation}
\label{av:Hamiltonian}
    H_{\CD}^2 = \frac{8 \pi G}{3} \average{\rho} -\frac{\average{\CR}+\CQ_{\CD}}{6} +\frac{\Lambda}{3}\;,
\end{equation}
where $\CQ_{\CD}$ denotes the backreaction and encapsulates the effects of the inhomogeneities on the domain's evolution in time and $H_{\CD}=\dot{a}_{\CD}/a_{\CD}$ is the domain-dependent Hubble parameter. For irrotational dust, the backreaction term is given explicitly by:
\begin{equation}
    \CQ_{\CD} = \frac{2}{3}\average{\left(\theta - \average{\theta}\right)^2} - 2\average{\sigma^2} \; .
\end{equation}

Equation (\ref{av:Hamiltonian}) describes the averaged Hamiltonian constraint on the given domain $\CD$, which has to hold for every hyper-surface of the constant proper time of the dust. Moreover, the domain-dependent Hubble parameter is related to the volume of the domain:
\begin{equation}
    H_{\CD} = \frac{\dot{V}_{\CD}}{V_{\CD}} \;,
\end{equation}
and can be used to derive the maximum volume. 
Due to the presence of the backreaction term, an additional assumption should be made to calculate a domain-dependent scale factor $a_{\CD}$. This will be obtained by applying the relativistic Zel'dovich approximation, in a similar fashion as in \cite{flat}.

\subsection{Relativistic Zel'dovich approximation}
The standard Zel'dovich approximation \cite{zeldovich} is an extrapolation scheme rooted in the Lagrangian perturbation theory. Zel'dovich's original idea was to assume a simple form of the fluid's trajectory $\mathbf{f}$, or in other words, a simple relation between the Lagrangian ($\bX$) and Eulerian ($\bx$) coordinates: 
\begin{equation}
    \bx = \mathbf{f}(\bX,t)= a(t)\left(\bX+b(t) P_0\right)\;,
\end{equation}
where $a(t)$ is the background scale factor, $P_0$ is the initial displacement related to the initial potential $P_0=\nabla \Phi_0$ and $b(t)$ is the time-dependent growth function. The following is an exact integral of the density field:
\begin{equation}
\label{ZeldovichNewton}
    \rho = \frac{\rho_0}{J} \;\;,\;\; J = \mathrm{det}\left(\frac{\partial f^i}{\partial X^j}\right) \;,
\end{equation}
where $J$ is the Jacobian of the transformation between the Eulerian and Lagrangian coordinates. Equation (\ref{ZeldovichNewton})
can be expanded around $b(t)\partial_{X}P_0 << 1$ and compared with the Eulerian density perturbation formula to obtain the explicit form of the time-dependent function $b(t)$. However, equation (\ref{ZeldovichNewton}) is intentionally not linearized and as such can probe the mildly non-linear regime of the structure formation. In other words,
the Zel'dovich approximation is a restricted first-order Lagrangian perturbation scheme (the initial acceleration is parallel to the velocity, and there is no decaying mode of density perturbation) and the associated extrapolation procedure is to keep the density field non-linear. In a similar spirit the relativistic version of the Zel'dovich approximation (RZA) was developed: \cite{RZA1}. The main idea  is to express Einstein's equations in the $3+1$ synchronous and co-moving setup exclusively in terms of spatial co-frames $\eta^a_{\;\;i}$, its functionals and derivatives. This leads to (see \cite{RZA1} for a detailed derivation):
\begin{eqnarray}
\label{eisnteinlag}
&{}&\delta_{ab} \ddot{\eta}^a_{\;\;[i}\eta^b_{\;\;j]} =0 \;\;,\;\;
\frac{1}{2}\epsilon_{abc}\epsilon^{ikl}\eta^a_{\;\;i}\eta^b_{\;\;k}\eta^c_{\;\;l}= \Lambda J-4\pi G J_0 \rho_0 \;\;, \nonumber \\
&{}&\left(\epsilon_{abc}\epsilon^{ikl}\dot{\eta}^a_{\;\;i}\eta^b_{\;\;k}\eta^c_{\;\;l}\right)_{|i} = \left(\epsilon_{abc}\epsilon^{ikl}\dot{\eta}^a_{\;\;i}\eta^b_{\;\;k}\eta^c_{\;\;l}\right)_{|j} \;\;, \nonumber \\
&{}&\epsilon_{abc}\epsilon^{mkl}\dot{\eta}^a_{\;\;m}\eta^b_{\;\;k}\eta^c_{\;\;l} = 16\pi G J_0 \rho_0+2\Lambda J -JR \nonumber \;\;, \\
&{}&\frac{1}{2}\left(\epsilon_{abc}\epsilon^{ikl}\ddot{\eta}^a_{\;\;i}\eta^b_{\;\;k}\eta^c_{\;\;l} -\frac{1}{3}\epsilon_{abc}\epsilon^{mkl}\ddot{\eta}^a_{\;\;m}\eta^b_{\;\;k}\eta^c_{\;\;l}\delta^i_{\;\;j}\right)+\nonumber \\ &{}&\left(\epsilon_{abc}\epsilon^{ikl}\dot{\eta}^a_{\;\;i}\dot{\eta}^b_{\;\;k}\eta^c_{\;\;l}-\frac{1}{3}\epsilon_{abc}\epsilon^{mkl}\dot{\eta}^a_{\;\;m}\dot{\eta}^b_{\;\;k}\eta^c_{\;\;l}\delta^i_{\;\;j}\right) \nonumber =-J\tau^i_{\;\;j} \;. 
\end{eqnarray}
The vertical bar stands for the covariant spatial derivative, the $0$ subscript denotes the initial value, the coframe indices start with $a$ and coordinate indices start with $i$, $J$ is the determinant of the transformation from the
coordinate to the non-coordinate basis and $\tau^i_{\;\;j}$ is the off-diagonal spatial Ricci tensor. 
The procedure is then to split the co-frames into  background (e.g. FLRW) and perturbation, linearize (\ref{eisnteinlag}) to find the explicit solution for the perturbed co-frames and then evaluate all functionals (e.g. density, spatial curvature, expansion) retaining all orders without any further truncation. The so-obtained co-frame field reads:
\begin{equation}
    \eta^a_{\;\;i} = a(t)\left(
    \delta^a_{\;\;i} +P^a_{\;\;i}(\bX,t) +\xi(t) \dot{P}^a_{\;\;i}(\bX,t)\right)\;,
\end{equation}
where $P^a_{\;\;i}$ is a deviation field and
\begin{equation}
    \xi(t) = \frac{q(t)-q(t_0)}{\dot{q}_0}{}\;\;,\;\;\ddot{q}+2\frac{\dot{a}}{a}\dot{q}+\left(3\frac{\ddot{a}}{a}-\Lambda\right)\left(q+q(t_0)\right)=0\; ,
\end{equation}
with $q$ denoting the growing mode. With this {\it ansatz} the scalar averaged equations become closed. 
\section{Results}
The Jacobian of the transformation from co-frames to coordinates can be expressed, using the relativistic Zel'dovich approximation, as:
\begin{equation}
    J  = a^2\CJ = a^3\left(1+\xi \inI_i +\xi^2 \inII_i + \xi^3 \inIII_i\right) \;,
\end{equation}
where the initial invariants of the extrinsic curvature tensor $\theta_{ij}=\frac{1}{2}\dot{g}_{ij}$ (with $g_{ij}$ denoting the spatial metric) are given by:
\begin{equation}
    \inI = \mathrm{tr}(\theta_{ij}) \;\;,\;\;\inII = \frac{1}{2}\left((\mathrm{tr}(\theta_{ij}))^2-\mathrm{tr}(\left(\theta_{ij}\right)^2)\right) \;\;,\;\;\inIII = \det (\theta_{ij}) \;,
\end{equation}
and $\CJ$ is called the peculiar volume deformation. We can decompose the domain-dependent Hubble function into the background and peculiar Hubble flow:
\begin{equation}
    H_{\CD} = H+\frac{1}{3}\frac{\average{\dot{\CJ}}}{\average{\CJ}}
\end{equation}
and impose the condition for the turnaround to be $H_{\CD} = 0$. This corresponds to the maximum volume that a given object can acquire during its evolution. Using the averaged acceleration equation:
\begin{equation}
   3 \frac{\ddot{a}_{\CD}}{a_{\CD}} +4\pi G \average{\rho} -\Lambda= \CQ_{\CD} \;,
\end{equation}
and the backreaction formula:
\begin{equation}
    \CQ_{\CD} = \frac{\aaverage{\ddot{\CJ}}}{\aaverage{\CJ}}-\frac{\ddot{\xi}}{\dot{\xi}}\frac{\aaverage{\dot{\CJ}}}{\aaverage{\CJ}}-\frac{2}{3}\left(\frac{\aaverage{\dot{\CJ}}}{\aaverage{\CJ}}\right)^2 \;\;,\;\;\aaverage{\CA} = \frac{1}{V_i}\int \CA \rmd^3\bX\;;\,
\end{equation}
where $V_i$ is the initial volume, together with the maximum volume condition we obtain an expression for the maximum volume attainable by cosmological structures
    \begin{equation}
      \label{volumemax}
V_{max} = \frac{M}{\rho_H\left(1+3H\left(\frac{\dot{q}}{q}\right)^{-1}\right)} \;\;,
      \end{equation}
where $\rho_H$ is the background density. Equation (\ref{volumemax}) depends on the mass of the structure and the background parameters and thus can be used to test different cosmological models.  

\section{Conclusions}

Independent cosmological tests are a necessary ingredient of the current research aiming at solving tensions in the $\Lambda$CDM model. With the help of relativistic Lagrangian perturbation theory and the scalar, averaged Einstein's equations we derived a formula for the maximum volume of the collapsing structure as a function of its mass and the background parameters.  The volume of collapsing structure can in principle be hard to measure unless we assume some approximate spatial symmetries e.g. spherical or axis-symmetry, however, the ongoing era of massive sky surveys will allow to map peculiar velocities in large-scale structures to unprecedented precision and thus make our proposed test realistically applicable.

The analysis presented in this text is based on two main pillars: the turnaround condition given by the vanishing of the averaged scalar expansion and the RZA. While RZA relies on the existence of a cosmological background and should not be applied beyond the mildly non-linear regime (reasonable conditions for the large-scale structure formation), our turnaround definition is general and can be used within other approaches, gravitational theories or simulations. In this context, we highlight two relativistic frameworks allowing for extension of the presented research, namely the silent universe approach with simplified initial conditions \cite{BolSilUni} and the generalized RZA (GRZA) \cite{GRZA}. The latter offers a metric-based approach and a direct connection with RZA, LTB and quasi-spherical Szekeres models of class I, being suitable for generalizing our results while retaining the methodology; the former provides a scenario for performing relativistic simulations and using them to test our result and look for potential generalizations.

An interesting idea, related to our analysis is that of the finite infinity (see e.g. \cite{Ellis}). The notion of spatial and null infinities plays a major role in many branches of general relativity. However, in a realistic universe, fundamental observers can be either within gravitationally bound structures or subject to the Hubble expansion. In this context, a definition of a boundary (finite infinity) that would be far enough from the object for the metric to be approximately flat and yet close enough not to be affected by the cosmic expansion or other structures is required. In \cite{wiltshire} an operational definition was proposed by Wiltshire who required for the volume within this boundary to be on average non-expanding. This is equivalent to our condition - putting the domain-dependent Hubble parameter to zero. Exploration of the possible connections between these two notions will be the subject of future investigation.
  
\section*{Acknowledgements}
JJO and IDG acknowledge the support of the National Science Centre (NCN, Poland) under the Sonata-15 research grant UMO-2019/35/D/ST9/00342.

\end{document}